\documentclass[twocolumn,showpacs,preprintnumbers,superscriptaddress,amsmath,amssymb,APSl,prd,nofootinbib]{revtex4-1}

\usepackage{dcolumn}
\usepackage{bm}
\usepackage{ifpdf}
\usepackage{hyperref}
\usepackage{dcolumn}
\usepackage{bm}
\usepackage[spanish,english]{babel}
\usepackage{amsfonts}
\usepackage{amssymb}
\usepackage{graphicx}
\usepackage[latin1]{inputenc}

\newcommand \be{\begin{equation}}
\newcommand \en{\end{equation}}
\newcommand \bea{\begin{eqnarray}}
\newcommand \ena{\end{eqnarray}}

\begin{document}

\title{Modified gravity in three dimensional metric-affine scenarios}

\author{Cosimo Bambi} 
\email{bambi@fudan.edu.cn}
\affiliation{Center for Field Theory and Particle Physics and Department of
Physics, Fudan University, 220 Handan Road, 200433 Shanghai, China}
\affiliation{Theoretical Astrophysics, Eberhard-Karls Universit\"at T\"ubingen, Auf der Morgenstelle 10, 72076 T\"ubingen, Germany}

\author{M. Ghasemi-Nodehi} 
\email{nmghasemi14@fudan.edu.cn}
\affiliation{Center for Field Theory and Particle Physics and Department of
Physics, Fudan University, 220 Handan Road, 200433 Shanghai, China}

\author{D. Rubiera-Garcia} 
\email{drubiera@fudan.edu.cn}
\affiliation{Center for Field Theory and Particle Physics and Department of
Physics, Fudan University, 220 Handan Road, 200433 Shanghai, China}


\date{\today}

\begin{abstract}
We consider metric-affine scenarios where a modified gravitational action
is sourced by electrovacuum fields in a three dimensional space-time. Such
scenarios are supported by the physics of crystalline structures with
microscopic defects and, in particular, those that can be effectively
treated as bi-dimensional (like graphene). We first study the case of
$f(R)$ theories, finding deviations near the center as compared to the
solutions of General Relativity. We then consider Born-Infeld gravity,
which has raised a lot of interest in the last few years regarding its
applications in astrophysics and cosmology, and show that new features
always arise at a finite distance from the center. Several properties of
the resulting space-times, in particular in presence of a cosmological
constant term, are discussed.
\end{abstract}

\maketitle

\section{Introduction}

One hundred years after the development of Einstein's theory of General Relativity (GR), many experimental results have confirmed its success \cite{ReviewGR}.
This theory, however, is still troubled by the presence of space-time singularities \cite{singularities}, where it breaks down and loses its reliability. In
this sense, there is the widespread belief that the solution to this problem will be achieved once GR and quantum theory merge into a single, unified framework
-a quantum theory of gravity-, like in string theory \cite{strings} or loop quantum gravity. Another approach to this problem consists of obtaining some
insights using classical theories of gravity including higher-order corrections in the curvature invariants \cite{Cembranos}. The simplest such example is that
of $f(R)$ theories, where $f$ is a given function of the curvature scalar, $R$. Indeed, such an approach has allowed to shape different astrophysical and
cosmological phenomena (see e.g. \cite{fapp} for a review).

Given the mathematical complexity of GR and its extensions, a useful way to study their physics relies on considering their three dimensional counterparts
\cite{Carlip}. Here analytic solutions can be obtained and characterized, which might provide useful insights for four dimensional black hole physics. Indeed,
three dimensional black holes may arise from effective descriptions of four dimensional gravity in string theory \cite{Witten}. In this simplified scenario, a
relevant finding is that of the BTZ solution \cite{BTZ} (see also \cite{BTZ2}), where a nonsingular Anti-de Sitter space-time arises as a bound state separated
from the continuous black hole spectrum by a mass gap. This solution and its extensions have been widely studied due to their potential implications within the
AdS/CFT correspondence \cite{AdS/CFT} and systems of condensed matter physics. Indeed, regular solutions sourced by an electromagnetic field date back to the
Bardeen solution \cite{Bardeen} and its nonlinear extensions \cite{AB}. In three space-time dimensions, regular black hole solutions have been obtained as
generalizations of the Bardeen solution \cite{Cataldo1}.  Other three dimensional space-times include wormhole solutions in nonlinear electrodynamics
\cite{Cataldo2,NED}, charged shells scenarios \cite{Habib1}, black holes with dilaton fields \cite{Yamazaki},  magnetic solutions \cite{Dias}, and further
extensions of the BTZ solution \cite{BTZe}.

When dealing with extensions of GR, however, one is immediately faced with the troubles of higher-order field equations and ghost-like instabilities. One is
then forced to constrain the parameters of the theory (when possible) in order to get rid of such undesirable features \cite{Lovelock}. However, another
approach to this problem comes from the study of crystalline structures and, in particular, Bravais crystals in solid state physics \cite{Kittel}. At the
microscopic level they are defined by a discrete net-like arrangement of atoms. However, it has been found that at the macroscopic level they admit a
description in terms of an effective differential geometry \cite{Kroner}. If the crystal contains no defects on its microscopical structure, then the
appropriate geometry is the standard Riemannian one of GR. However, all crystals contain defects of different sorts, and the Riemannian description no longer
holds. A metric-affine geometry becomes necessary \cite{Kroner2}. If such defects are point-like then one has non-metricity \cite{Falk81} (the line-like defects
called dislocations require instead Cartan's torsion \cite{Bilby}), which implies that metric and affine connection are not constrained by the standard
compatibility condition of GR (i.e., the latter is not given by the Christoffel symbols of the former), but are instead independent entities (see \cite{Zanelli}
for a pedagogical discussion). Independent variations of the action are then performed with respect to the metric and the connection, and the specific form of
the latter is determined through the corresponding equation \cite{olmo-review}. It turns out that this approach leads to second-order equations and absence of
dynamical instabilities for a large family of gravitational theories.

The application of this framework to a three dimensional space-time could be of great relevance regarding some systems of solid state physics like graphene,
which consists of an extremely thin layer of hexagonal atoms \cite{graphene}, so it can be effectively treated as bi-dimensional. The presence of defects on its
microstructure raises the interest in the consideration of metric-affine theories. Let us point out, however, that these geometrical scenarios have received
little attention so far and, consequently, they are still poorly understood. The goal of this paper is to made a modest step in this direction, and consider
metric-affine gravities with non-metricity in a three dimensional space-time. We focus on two classes of theories, $f(R)$ and Born-Infeld gravity, the latter
being supported by some results of solid state physics systems with defects. The dynamics of the new gravitational system is excited by considering an
electromagnetic (Maxwell) field.

\section{$f(R)$ theories in metric-affine approach}

The action of our theory is given by

\be \label{eq:action}
S=\frac{1}{2\kappa^2} \int d^3x \sqrt{-g} f(R)+ S_m(g_{\mu\nu},\psi_m),
\en
where $\kappa^2$ is Newton's constant in suitable dimensions, $g$ is the determinant of the space-time metric $g_{\mu\nu}$, $f(R)$ is a given function
(unspecified at this stage) of the curvature scalar $R=g_{\mu\nu}R^{\mu\nu}$ with the definition $R_{\mu\nu} \equiv {R^\alpha}_{\mu\alpha\nu}$, where

\be
{R^\alpha}_{\beta\mu\nu}=\partial_{\mu}
\Gamma^{\alpha}_{\nu\beta}-\partial_{\nu}
\Gamma^{\alpha}_{\mu\beta}+\Gamma^{\alpha}_{\mu\lambda}\Gamma^{\lambda}_{\nu\beta}-\Gamma^{\alpha}_{\nu\lambda}\Gamma^{\lambda}_{\mu\beta}
\en
is the Riemann tensor. $S_m$ is the matter action and it depends on the matter fields, denoted collectively as $\psi_m$, and is assumed to only couple through
the metric. In these definitions, the connection $\Gamma^{\lambda}_{\mu\nu}$ is a priori independent of the metric (metric-affine or Palatini approach). In
addition, we shall assume vanishing torsion, $T^{\lambda}_{\mu\nu} \equiv \Gamma^{\lambda}_{\mu\nu}- \Gamma^{\lambda}_{\nu\mu}=0$ \cite{Torsion}. As we will be
interested in electrovacuum space-times, and to work as general as possible, the matter sector of our theory is that of nonlinear electrodynamics, with action

\be \label{eq:matter}
S_m= \int d^3x \sqrt{-g} \varphi(X),
\en
where $\varphi(X)$ is a given function of the field invariant $X=-\frac{1}{2}F_{\mu\nu}F^{\mu\nu}$ constructed with the field strength tensor
$F_{\mu\nu}=\partial_{\mu}A_{\nu} - \partial_{\nu}A_{\mu}$, with $A_{\mu}$ the vector potential. To obtain the field equations for the theory defined by
Eqs.(\ref{eq:action}) and (\ref{eq:matter}) with the constraints above, we perform independent variations with respect to the metric and the connection, which
yields (the details of this variation can be found in \cite{f(R)5D} for arbitrary dimension, and thus we only bring here the final result)

\bea
f_R R_{\mu\nu} - \frac{f}{2}g_{\mu\nu}&=&\kappa^2 T_{\mu\nu}, \label{eq:metric}\\
\nabla_{\lambda}^{\Gamma} (\sqrt{-g} f_R g^{\mu\nu})&=&0, \label{eq:connection}
\ena
where we have introduced the short-hand notation $f_R \equiv df/dR$, and $T_{\mu\nu}=-\frac{2}{\sqrt{-g}} \frac{\delta S_m}{\delta g^{\mu\nu}}$ is the
energy-momentum tensor of the matter, which is obtained as

\be \label{eq:Tmununed}
T_{\mu\nu}=2\varphi_X F_{\mu\alpha}{F^\alpha}_{\nu}- \varphi(X) g_{\mu\nu},
\en
where $\varphi_X \equiv d\varphi/dX$. Non-metricity appears explicitly in this framework through Eq.(\ref{eq:connection}). Before solving these equations, let
us first take a trace in (\ref{eq:metric}), which yields

\be \label{eq:trace}
Rf_R-\frac{3}{2} f=\kappa^2 T,
\en
where $T =2\varphi - 4X\varphi_X$ is the trace of $T_{\mu\nu}$. Note that (\ref{eq:trace}) is not a differential equation but just an algebraic relation
$R=R(T)$ extending the GR one, $R=-\kappa^2 T$, to the case of metric-affine $f(R)$ gravity coupled to nonlinear electrodynamics.  This implies that $f(R)$ is
just a function of the matter fields.

Now, to solve the system of field equations (\ref{eq:metric}) and (\ref{eq:connection}) we take advantage of the fact that (\ref{eq:connection}) can be formally
rewritten as the standard metric-compatibility condition, namely

\be
\nabla_{\lambda}^{\Gamma}(\sqrt{-h}h^{\mu\nu})=0,
\en
which means that $\Gamma_{\mu\nu}^{\lambda}$ is the Levi-Civita connection of a rank-two tensor $h_{\mu\nu}$. A little algebra shows that the relation between
$g_{\mu\nu}$ and $h_{\mu\nu}$ is given by the conformal transformation

\be \label{eq:h-g}
h_{\mu\nu}=f_R^2 g_{\mu\nu} \hspace{0.1cm} ; \hspace{0.1cm}  h^{\mu\nu}=f_R^{-2} g^{\mu\nu} .
\en
In terms of $h_{\mu\nu}$ the metric field equations (\ref{eq:metric}) admit the following simple representation

\be \label{eq:Rmunu}
{R_\mu}^{\nu}(h)=\frac{1}{f_R^3} \left(\frac{f}{2} {\delta_\mu}^{\nu} +\kappa^2 {T_\mu}^{\nu} \right).
\en
This is a set of Einstein-like second-order field equations for the metric $h_{\mu\nu}$, with the right-hand-side only depending on the matter fields. The field
equations for $g_{\mu\nu}$ follow from those of $h_{\mu\nu}$ through the matter-dependent conformal transformation of Eq.(\ref{eq:h-g}) and, therefore, they are
second-order as well. In vacuum, ${T_\mu}^{\nu}=0$, the field equations boil down to those of GR with possibly a cosmological constant term, depending on the
particular $f(R)$ function chosen. In summary, the system of equations (\ref{eq:Rmunu}), with the relations (\ref{eq:Tmununed}), (\ref{eq:trace}) and
(\ref{eq:h-g}) provide a solution for a given problem once the forms of $f(R)$ and $\varphi(X)$ are specified.

We note that the field equations (\ref{eq:Rmunu}) are in sharp contrast with the more standard case of the metric formulation of $f(R)$ gravity (see e.g.
\cite{ReviewfR,Review} for reviews of the theory and confrontation between these two approaches), where they take the form

\begin{equation}
f_R R_{\mu\nu} -\frac{f}{2} g_{\mu\nu} -\nabla_{\mu}\nabla_{\nu} f_R + g_{\mu\nu} \Box f_R=\kappa^2 T_{\mu\nu}
\end{equation}
This is a set of fourth-order differential field equations, whose resolution
is, in general, a non-trivial task. In this sense, simplifications such as
that of constant curvature, $R=R_0$, which allows to remove the contribution
of the fourth-order terms, allows for the obtention of explicit solutions
(see e.g. \cite{Review}). In our case, the presence of non-metricity,
$Q_{\mu\nu\lambda} \equiv \nabla_{\mu} g_{\nu\lambda}$, which measures the
failure of the independent connection to be metric with respect to
$g_{\mu\nu}$ \cite{Sotiriou}, introduces a relative (conformal) deformation
between the Riemannian structure associated to $h_{\mu\nu}$, and that
associated to the physical metric $g_{\mu\nu}$. Note that, because of the
different structure of the field equations and methodology followed to find
analytical solutions in the metric and metric-affine approaches, here we will
be mostly interested on comparing the deviations with respect to GR.

\subsection{Electrovacuum solutions}

We are interested in solving the field equations for a three dimensional, static, spherically symmetric line element of the form

\begin{equation}
ds^2=g_{tt}dt^2 +g_{rr}dr^2 +r^2 d\theta^2.
\end{equation}
For a purely electric field the only-nonvanishing component of the field strength tensor in this setting is $F^{tr} \neq 0$. From the matter field equations,
$\delta S_m/ \delta A_{\nu}=0 \rightarrow \nabla_{\mu}(\sqrt{-g} \varphi_X F^{\mu\nu})=0$, it follows that this component satisfies

\be
F^{tr}=\frac{q}{r\sqrt{-g_{tt}g_{rr}} \varphi_X},
\en
where $q$ is an integration constant identified as the electric charge associated to a given solution. To formulate the matter side of the field equations
(\ref{eq:Rmunu}) in a way independent of the metric, we note that in terms of $X=q^2/(r^2 \varphi_X^2)$ the energy-momentum tensor reads

\bea \label{eq:Tmunu}
{T_\mu}^{\nu}=
\left(
\begin{array}{cc}
(\varphi-2X\varphi_X) \hat{I}_{2\times 2}&  \hat{0}_{1 \times 2} \\
\hat{0}_{2 \times 1} & \varphi  \\
\end{array}
\right). \label{eq:em}
\ena

To proceed further, we need to specify the choice for $f(R)$ and $\varphi(X)$. For the gravity part, we consider the theory

\be
f(R)=R+\alpha R^{3/2}
\en
by computational convenience, since the trace equation (\ref{eq:trace}) yields the linear relation $R=-2\kappa^2 T$, thus simplifying the analytic resolution
of the field equations. In this theory $\alpha$ is some constant (with dimensions), assumed to be small, which controls the deviation with respect to GR
solutions, to which the new configurations will be compared. For the matter section, we take the choice $\varphi(X)=X/(4\pi)$, corresponding to Maxwell
electrodynamics, which in three dimensions yields a non-vanishing trace $T=q^2/(4\pi r^2)$ (in four dimensions, however, the Maxwell energy-momentum tensor is
traceless and one needs to use a nonlinear theory of electrodynamics like the Born-Infeld one in order to obtain modified dynamics \cite{or11}). Thus we have
$R=2r_q^2/r^2$, where $r_q^2=\kappa^2 q^2/(4\pi)$ is a charge scale. Next we introduce two line elements, one for the metric $h_{\mu\nu}$ as

\be \label{eq:ansatz1}
d\tilde{s}^2=-A(x)e^{2\xi(x)}dt^2+\frac{1}{A(x)} dx^2 + \tilde{r}^2(x)d\theta^2,
\en
and another for $g_{\mu\nu}$ as

\be \label{eq:ansatz2}
ds^2=-B(x) dt^2 + C(x)dx^2 + r^2(x)d\theta^2,
\en
where $r^2(x)$ and $\tilde{r}^2(x)$ are, in general, functions of the coordinate $x$, while $\xi(x)$ and $A(x)$ are functions to be determined by solving the
field equations and the functions $B(x)$ and $C(x)$ follow by consistency from the relations (\ref{eq:h-g}). Inserting (\ref{eq:ansatz1}) into (\ref{eq:Rmunu}),
and taking into account the matter symmetry, $T_t^t=T_x^x$, from the combination ${R_t}^t={R_x}^x$ it follows that $\xi(x)=$constant. Through some
redefinitions, and without loss of generality, the line element (\ref{eq:ansatz1}) can be rewritten as

\be \label{eq:ansatz}
d\tilde{s}^2=-A(x)dt^2+\frac{1}{A(x)} dx^2 + x^2 d\theta^2.
\en
Introducing the standard mass ansatz in three dimensions, $A(x)=-M(x)$, from the component ${(_\theta}^{\theta})$ of the field equations (\ref{eq:Rmunu}) one
finds

\be
M_x=2\frac{x}{f_R^3} \left(\frac{r_q^2}{r^2} \right)  \left[1+\frac{\alpha}{\sqrt{2}} \frac{r_q}{r}  \right].
\en
The relation between the coordinates $x$ and $r(x)$ also follows from Eqs.(\ref{eq:h-g}) as $x=f_R r(x)$, which simply implies a shift in the radial coordinate
so $dx/dr=1$. Let us first assume $\alpha>0$. Introducing a new variable $z \equiv \sqrt{2}r /(3\alpha r_q) $, we perform the integration of the mass function
as $M=M_0 + 2r_q^2G(z)$
where the function $G(z)$ satisfies:

\be \label{eq:Gz1}
\frac{dG}{dz}=\frac{1}{z} \frac{\left(1+\frac{1}{3z}\right)}{\left(1+\frac{1}{z}\right)^2}.
\en
Using again Eq.(\ref{eq:h-g}), we find $B(x)=A(x)/f_R^2$ and, therefore, a full solution to this problem is achieved. For large distances, we get $f_R \sim 1$
and $dG/dz \sim 1/z$, which upon integration yields $G(z)\sim \log(z)$, and the standard solution found in GR, $A(x)=-M_0-2r_q^2 \log[z]$, is recovered
\cite{Cataldo1}. Note that this asymptotic behaviour is neither flat nor (Anti-)de Sitter, which is a generic feature of any such solutions in three dimensions
in absence of cosmological constant term

The integration of (\ref{eq:Gz1}) is analytical, and can be expressed as

\be
G(z)=\frac{1}{3} \left(\frac{2}{1+z} +3\log[1+z]\right).
\en
As one gets close to $z=0$, deviations with respect to the GR solution are found. Going back to the coordinate $r$ to compare with the GR case, the function
$A(r)/f_R^2$ can be expanded in series around $r=0$ (defining a new constant $\tilde{\alpha}\equiv 3\alpha/\sqrt{2} \neq 0$) as

\begin{figure}[t]
\includegraphics[width=0.5\textwidth]{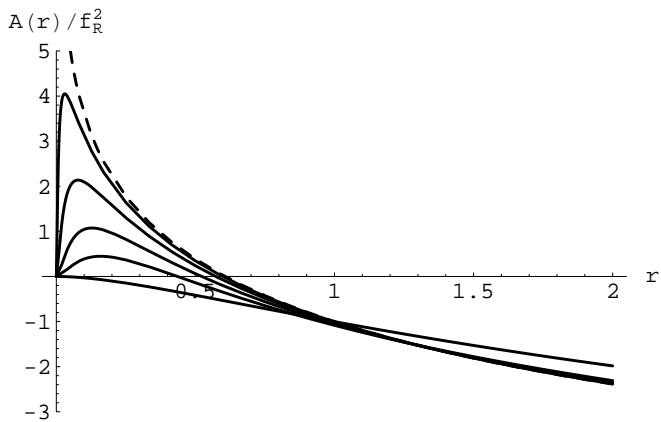}
\caption{Behaviour of the metric function $A(r)/f_R^2$ for $f(R)$ gravity with a Maxwell field ($\tilde{\alpha} >0$). In this figure, $M_0=r_q=1$. From the GR
solution ($\tilde{\alpha}=0$, dashed curve), the set of solid curves corresponds to growing values of $\tilde{\alpha}=0.005,0.02,0.05,0.1,0.5$. They approach
the (non-asymptotically flat) GR curve for large $r$, $A(r) \simeq -M_0-2r_q^2 \log[r]  + b \tilde{\alpha} /r + O(\tilde{\alpha}^2)$, with $b(M_0,r_q)$ a
constant. Note that at $r=0$ all solutions with $\tilde{\alpha} \neq 0$ go to zero. For $\tilde{\alpha}>\tilde{\alpha}_c(M=r_q=1) \simeq 0.3$ the horizon
disappears. \label{fig:1}}
\end{figure}

\be
 A(r)/f_R^2 \simeq -\left(\frac{4}{3\tilde{\alpha}^2} +\frac{M_0}{r_q^2 \tilde{\alpha}^2} +\frac{2\log[\tilde{\alpha} r_q]}{\tilde{\alpha}^2} \right) r^2 + O
 \left( r^3 \right)
\en
This expansion implies that, for any value of $\tilde{\alpha} \neq 0$, at the center all the solutions approach $A(r)/f_R^2 \rightarrow 0$ (as compared to
$A^{GR} \rightarrow + \infty$ in the GR case) due to the factor $1/f_R^2$, which goes to $0$ due to the unboundness of the Maxwell field. In terms of horizons,
a numerical analysis determines that, depending on the values of $M_0$ and $r_q$, there will be a single horizon (like in GR) or none (see Fig.\ref{fig:1}). The
horizon disappears for a fixed value of $r_q$, when $\tilde{\alpha}>\tilde{\alpha}_c(r_q)$, where $\tilde{\alpha}_c(r_q)$ is a critical value that grows with
$r_q$. This dramatic change in the behaviour of the metric at the center for any $\tilde{\alpha} \neq 0$  translates into a worsening of the degree of
divergence of the curvature invariants. For example, the Kretchsman scalar, $K \equiv {R^\alpha}_{\beta \gamma \delta} {R_\alpha}^{\beta \gamma \delta}$,
behaves as $\sim r_q^4/r^4$ in the GR case, but as $\sim 1/r^8$ in the $f(R)$ case. This space-time seems to be quite pathological.

\begin{figure}[t]
\includegraphics[width=0.5\textwidth]{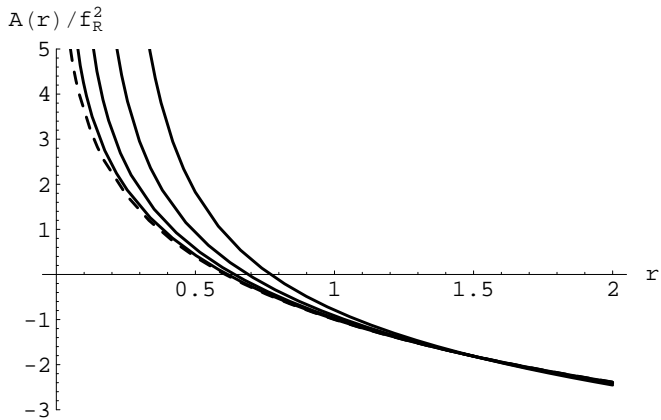}
\caption{Behaviour of the metric function $A(r)/f_R^2$ for $f(R)$ gravity with a Maxwell field ($\hat{\alpha} =-\tilde{\alpha}>0$). In this figure, $M_0=r_q=1$.
From the GR solution ($\hat{\alpha}=0$, dashed curve), the set of solid curves corresponds to growing values of $\hat{\alpha}=0.005,0.02,0.05,0.1$, and approach
the (non-asymptotically flat) GR curve for large $r$, having a horizon always. Note that for any $\hat{\alpha} \neq 0$, no solution reaches $r=0$, but instead
they are only defined beyond $r>r_q \hat{\alpha}$. \label{fig:2}}
\end{figure}

Let us consider now the case with $\hat{\alpha}=-\tilde{\alpha}>0$. In terms of a variable $z=r/(\hat{\alpha}r_q)$, we now find $G_z=\frac{1}{z}
\frac{\left(1-\frac{1}{3z}\right)}{\left(1-\frac{1}{z}\right)^2}$, whose integration yields

\be
G(z)= \frac{1}{3} \left(\frac{2}{(z-1)} + 3\log[z-1] \right).
\en
This function ceases to be defined at $z=1$. Therefore, the metric function $A(r)/f_R^2$ is only defined for $r>r_q \hat{\alpha}$, and behaves around this
region as

\be
A(r)/f_R^2 \simeq \frac{4(r_q^5 \hat{\alpha}^3)}{3(r-r_q \hat{\alpha})^3} + O\left(\frac{1}{(r-r_q \hat{\alpha})^2} \right),
\en
which means that for any $\hat{\alpha} \neq 0$ it goes to $+ \infty$ as $ r \rightarrow r_q  \hat{\alpha} $, which is the same behaviour as the GR solutions. In
this case the strength of the central divergence is softened, from $K =12r_q^2/r^4$ in the GR case, to the dominant term $K \simeq 64r_q^2/(\hat{\alpha}^2
(r-r_q \hat{\alpha}))^2$. A horizon exists in all cases, and all solutions approach the GR one for $r \rightarrow \infty$ (see Fig.\ref{fig:2}). These results
are quite similar to the behaviour of solutions found in four dimensional metric-affine $f(R)$ gravity with a nonlinear electromagnetic field \cite{or11}.

Despite the different methodologies pursued in the metric-affine and metric approaches for $f(R)$ gravity mentioned above, let us mention that four dimensional
spherically symmetric space-times have been largely investigated in the latter, including analytical solutions of $f(R)=R^2$ \cite{SSS1},
Reissner-Nordstr\"om-like solutions with a de Sitter center in $f(R)=R+aR^2$ with non-linear electrodynamics \cite{SSS2}, perfect fluid solutions \cite{SSS3},
spherically symmetric stars and Schwarzschid de Sitter solutions in $f(R)=R-\gamma/R$  \cite{SSS4,SSS5}, and constant curvature solutions in several $f(R)$
models \cite{SSS6}.

\section{Born-Infeld gravity}

We turn now our attention to an extension of GR that has attracted a great deal of attention in the last few years, dubbed as Born-Infeld gravity \cite{BIgrav},
with many applications in astrophysical and cosmological scenarios \cite{BIapp}. For conceptual and technical simplicity, we can conveniently express its action
as

\be \label{eq:actionBI2}
S=\frac{1}{\kappa^2 \epsilon} \int d^3x \left[\sqrt{-q} - \lambda \sqrt{-g} \right]
\en
where $q$ is the determinant of the new metric $q_{\mu\nu} \equiv g_{\mu\nu} + \epsilon R_{\mu\nu} (\Gamma)$, and $\epsilon$ is a small parameter with
dimensions of length squared, which controls the deviation from GR. The meaning of $\lambda$ follows from a series expansion in $\epsilon$ as

\bea
\lim_{\epsilon \rightarrow 0} S&=&\frac{1}{2\kappa^2} \int d^3x \sqrt{-g} [R-2\Lambda]  \\
&-&\frac{\epsilon}{2\kappa^2} \int d^3x \sqrt{-g} \left(-\frac{R}{2} + R_{\mu\nu}R^{\mu\nu} \right) + O(\epsilon^2)\nonumber
\ena
where the first term corresponds to GR with a cosmological constant term $\Lambda=\frac{\lambda^2 -1}{\epsilon}$, and the second term are quadratic corrections.
This just a particular case of a family of Lagrangians that can be written under the form $f(R,R_{\mu\nu}R^{\mu\nu})$. Let us stress that, when formulated in
the metric-affine approach, its field equations are also second-order and ghost-free, as has been shown in four-dimensional settings \cite{or12}. It is also
worth pointing out that a three dimensional model of this $f(R,R_{\mu\nu}R^{\mu\nu})$ form has been found a few years ago by Bergshoeff, Hohm and Townsend, with
the interesting property that its formulation in the standard metric approach is ghost-free at the tree level \cite{bht}.

The meaning of the new metric $q_{\mu\nu}$ follows from the implementation of the metric-affine approach, so independent variations of the action
(\ref{eq:actionBI2}) with respect to the metric $g_{\mu\nu}$ and the connection $\Gamma^{\lambda}_{\mu\nu}$ yields the system of equations

\bea
\frac{\sqrt{-q}}{\sqrt{-g}} q^{\mu\nu} - \lambda g^{\mu\nu}&=&-\kappa^2 \epsilon T^{\mu\nu} \label{eq:metric2} \\
\nabla_{\lambda}^{\Gamma} (\sqrt{-q} q^{\mu\nu})&=&0 \label{eq:connection2}
\ena
so the independent connection is compatible with the metric $q_{\mu\nu}$ (but not with the space-time metric $g_{\mu\nu}$, since
$\nabla_{\lambda}^{\Gamma}(\sqrt{-g} g^{\mu\nu}) \neq 0$). The Born-Infeld gravity action (\ref{eq:actionBI2}) can be seen as the comparison of two dimensional
volume elements between the Riemannian metric $q_{\mu\nu}$ and the non-Riemannian one $\lambda g_{\mu\nu}$. As shown by Katanaev and Volovich \cite{Volovich},
such a comparison provides a natural measurement of the mass of point defects in crystalline structures, which is computed according to the difference of (two
dimensional) volume between a defected crystal (as described by $g_{\mu\nu}$) and an idealized perfect crystal without defects (defined by $q_{\mu\nu}$). Since
point defects break metricity \cite{Kroner2}, the action (\ref{eq:actionBI2}) in metric-affine approach with non-metricity seems to be favoured by the physics
of crystalline structures.

From Eq.(\ref{eq:connection2}) we see that the independent connection $\Gamma_{\mu\nu}^{\lambda}$ is given by the Christoffel symbols of $q_{\mu\nu}$ which,
therefore, carries the same meaning as $h_{\mu\nu}$ in the $f(R)$ case. Now the relation between $g_{\mu\nu}$ and $q_{\mu\nu}$ can also be expressed through the
algebraic transformation

\be \label{eq:gq}
q_{\mu\nu}=   \vert \hat{\Upsilon}\vert {(\Upsilon^{-1})_\mu}^{\alpha} g_{\alpha\nu} \hspace{0.1cm};\hspace{0.1cm}
q^{\mu\nu}=  \frac{1}{ \vert \hat{\Upsilon}\vert}  g^{\mu \alpha} {\Upsilon_\alpha}^{\nu} \ ,
\en
where we have defined a new object, $\hat{\Upsilon} \equiv \vert \hat{\Omega} \vert^{1/2} \hat{\Omega}^{-1}$, with the components of the matrix $\hat{\Omega}$
given by ${\Omega^\alpha}_{\nu} \equiv g^{\alpha \beta} q_{\beta \nu}$.  In terms of this matrix, the Born-Infeld Lagrangian in Eq.(\ref{eq:actionBI2}) is
simply expressed as $f_{BI}=\vert \hat{\Omega} \vert^{1/2}$ \cite{oor} and the metric field equations become

\be \label{eq:Omegadef}
\vert \hat{\Omega} \vert^{1/2} {(\Omega^{-1})^\mu}_{\nu}=\lambda {\delta^\mu}_\nu -\epsilon \kappa^2 {T^\mu}_\nu
\en
so once the energy-momentum tensor is specified, $\hat{\Omega}$ can be fully determined. In addition, this equation implies that $\hat{\Omega}$ only depends on
the matter energy-momentum tensor, ${T_\mu}^{\nu}$.

From the definition $\epsilon R_{\mu\nu}(\Gamma)=q_{\mu\nu}-g_{\mu\nu}$, we raise an index with the metric $q^{\alpha \nu}$ and use the above equation to write
the field equations in terms of $q^{\mu\nu}$ as

\be \label{eq:RmunuBI}
{R_\mu}^{\nu}(q)=\frac{\kappa^2}{|\hat\Upsilon|} \left[L_{BI} {\delta_\mu}^{\nu}+ {T_\mu}^{\nu} \right],
\en
where the Born-Infeld gravity Lagrangian is expressed as
\be \label{eq:LBI}
L_{BI}=\frac{|\hat\Upsilon|-\lambda}{\kappa^2 \epsilon} \ .
\en
In summary, Born-Infeld gravity admits a similar representation as $f(R)$ theories in three space-time dimensions. The relation between $g_{\mu\nu}$ and
$q_{\mu\nu}$ is not conformal, but still governed by algebraic transformations only depending on the matter-energy sources, ${T_\mu}^{\nu}$. In addition, the
same considerations regarding the second-order field equations and the recovery of GR in vacuum holds also in this case.

For an electromagnetic field (\ref{eq:Tmunu}), the equation (\ref{eq:Omegadef}) can be solved by introducing the ansatz

\be
\hat{\Omega}=
\left(
\begin{array}{cc}
\Omega_{+} \hat{I}_{2\times2} &  \hat{0}\\
\hat{0} &  \Omega_{-}\\
\end{array}
\right),
\en
where $I_{2 \times 2}$ and $\hat{0}$ are the identity and zero matrices, respectively, and we have defined the objects $\Omega_{-}=(\lambda + \hat{X})^2$,
$\Omega_{+}=(\lambda-\hat{X})(\lambda + \hat{X})$ with $\hat{X}=\frac{\epsilon \kappa^2}{4\pi} X$. The field equations (\ref{eq:RmunuBI}) in this case take the
form

\be \label{eq:Rmunuex}
\epsilon {R_\mu}^{\nu}(q)=
\left(
\begin{array}{cc}
\left(\frac{\Omega_{+}-1}{\Omega_{+}} \right) \hat{I}_{2\times2} &  \hat{0}\\
\hat{0} &  \left(\frac{\Omega_{-}-1}{\Omega_{-}} \right)\\
\end{array}
\right),
\en
As in the $f(R)$ case, the source symmetry ${T_t}^t={T_x}^x$ allows to write a metric in the form (\ref{eq:ansatz1}), and the mass ansatz in three space-time
dimensions, $A(x)=-M(x)$, yields a single independent equation to be solved:

\be \label{eq:MxBI}
\epsilon M_x=x  \frac{\left(\Omega_{-}-1 \right)}{\Omega_{-}}.
\en
From (\ref{eq:gq}) and the equations above, it follows that the relation between coordinates in the $q_{\mu\nu}$ and $g_{\mu\nu}$ geometries is given by
$x^2=r^2 \Omega_{-}$, which implies $dx/dr=\Omega_{+}/\Omega_{-}^{1/2}$. Plugging this result into (\ref{eq:MxBI}) we obtain

\be
\epsilon M_r=\frac{r (\Omega_{-}-1) \Omega_{+}}{\Omega_{-}}.
\en
Let us now introduce the length scale $\epsilon= l_{\epsilon}^2$ and the charge scale $r_q^2=\kappa^2 q^2/(4\pi)$ which, together with the space-time mass,
$M_0$, characterize the function $A(x)=-(M_0 + G(r)/l_{\epsilon}^2)$. The function $G(r)$ satisfies

\be
\frac{dG}{dr}=-\frac{r\left(\lambda-\frac{l_{\epsilon}^2 r_q^2}{r^2} \right) \left(\left(\lambda + \frac{l_{\epsilon}^2 r_q^2}{r^2} \right)^2-1\right)}{\lambda
+ \frac{l_{\epsilon}^2 r_q^2}{r^2}}
\en
and admits an analytic integration, namely

\be \label{eq:Gr}
G(r)=-\frac{1}{2} \left(\frac{l_{\epsilon}^4 r_q^4}{r^2}+ r^2(\lambda^2-1) + 2\frac{l_{\epsilon}^2 r_q^2 \log[l_{\epsilon}^2r_q^2 + \lambda r^2]}{\lambda}
\right),
\en
which completely determines the metric function $B(r)= A/\Omega_{+}$. The expansion of this function for $r \rightarrow \infty$ yields

\be \label{eq:asymp}
\frac{A}{\Omega_{+}} \simeq \frac{1-\lambda^2}{2l_{\epsilon}^2\lambda^2} r^2 - \frac{\lambda M_0 + r_q^2 (2\log[r]  + \log[\lambda])}{\lambda^3} +
O\left(\frac{1}{r} \right)^2
\en
which means that we find asymptotically de Sitter solutions if $l_{\epsilon}^2>0$ and $\lambda>1$, or if $l_{\epsilon}^2<0$ and $\lambda<1$, while
asymptotically Anti-de Sitter solutions are obtained when $l_{\epsilon}^2>0$ and $\lambda<1$, or $l_{\epsilon}^2<0$ and $\lambda>1$. When $\lambda=1$, the
asymptotic behaviour is neither (Anti-)de Sitter nor flat, but the standard three dimensional GR behaviour, namely, $ A(r)= - M_0 -2r_q^2 \log[r]$.

\begin{figure}[t]
\includegraphics[width=0.5\textwidth]{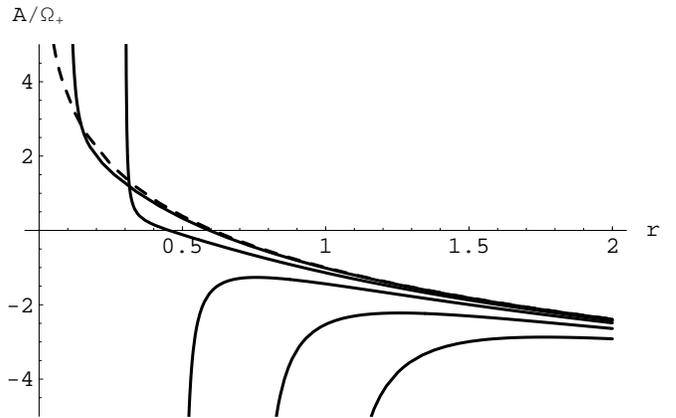}
\caption{Behaviour of metric function $A/\Omega_+$ for Born-Infeld gravity with Maxwell field with $\lambda=1$, parameters $M_0=r_q=1$, and length scale
$l_{\epsilon}^2>0$. Starting with GR solution ($l_{\epsilon}^2=0$, dashed curve) we increase $l_{\epsilon}^2=0.1,0.3,0.5,0.75,1$ (solid curves), which makes the
radius of the core to increase as well. A change in the behaviour occurs at $l_{\epsilon}^2=l_{\epsilon_c}^{2} \simeq 0.334$. Those solutions with
$l_{\epsilon}^2<l_{\epsilon_c}^{2}$ always have a horizon, while those with $l_{\epsilon}>l_{\epsilon_c}^{2}$ do not. All solutions converge asymptotically to
the GR case. \label{fig:3}}
\end{figure}

To study the behaviour around the center of the solutions let us first focus on the case $l_{\epsilon}^2>0$. We note that, given the fact that $\Omega_{+}$
vanishes at a non-vanishing (core) radius $r=l_{\epsilon} r_q/\lambda^{1/2}$, the series expansion around it yields the dominant term

\be
\frac{A}{\Omega_{+}} \simeq -\frac{l_{\epsilon} r_q(2M_0 \lambda + r_q^2(2\lambda^2-1+ 2\log[2l_{\epsilon}^2 r_q^2]))}{8\lambda^{7/2}
\left(r-\frac{l_{\epsilon}r_q}{\lambda^{1/2}} \right)} + \ldots
\en
This provides an involved interplay between the mass and charge parameters characterizing a solution, $M_0$ and $r_q$, the constant characterizing the
far-distance behaviour, $\lambda$, and the length scale $l_{\epsilon}^2$. To compare our results with those of GR, let us focus on the case $\lambda=1$. Given
the presence in this expansion of the term $\log[2l_{\epsilon}^2 r_q^2]$, we expect a change in the behavior for a sufficiently large value of $l_{\epsilon}^2$.
Numerical investigation of this question is plotted in Fig.\ref{fig:3}. The behaviour at the center depends on a critical value
$l_{\epsilon}^2=l_{\epsilon_c}^2$ (which depends on the specific amounts of mass and charge), in such a way that those solutions with
$l_{\epsilon}^2<l_{\epsilon_c}^2$ have a single cosmological horizon (like in the case of GR), while those satisfying $l_{\epsilon}^2>l_{\epsilon_c}^2$ do not,
though all cases asymptotically approach the GR solution, as already mentioned.

\begin{figure}[t]
\includegraphics[width=0.5\textwidth]{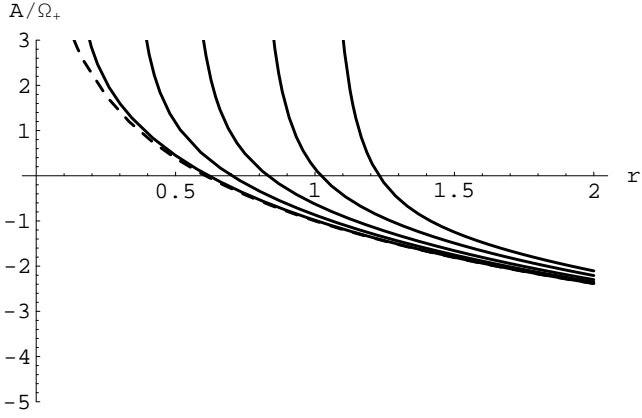}
\caption{Behaviour of metric component $A/\Omega_+$ for Born-Infeld gravity with Maxwell field with $\lambda=1$, parameters $M_0=r_q=1$, and length scale
$\tilde{l}_{\epsilon}^2\equiv -l_{\epsilon}^2>0$. Starting with GR solution ($\tilde{l}_{\epsilon}^2=0$, dashed curve) we increase
$\tilde{l}_{\epsilon}^2=0.1,0.3,0.5,0.75,1$ (solid curves), which makes the radius of the core to increase as well. All solutions converge asymptotically to the
GR case. \label{fig:4}}
\end{figure}

Let us consider now the behaviour at the center for those solutions with $\tilde{l}_{\epsilon}^2 \equiv - l_{\epsilon}^2 >0$. Now the expansion around
$r=\tilde{l}_{\epsilon} r_q/\lambda^{1/2}$ becomes

\bea
\frac{A}{\Omega_{+}} &\simeq & \frac{\tilde{l}_{\epsilon} r_q(2M_0 \lambda + r_q^2( 1 - 2 \lambda^2 + 2 \log[r-\frac{\tilde{l}_{\epsilon} r_q}{\lambda^{1/2}}]
))}{8\lambda^{7/2} \left( r - \frac{\tilde{l}_{\epsilon} r_q}{\lambda^{1/2}} \right)} \nonumber \\
&+& \frac{2\tilde{l}_{\epsilon} r_q \log[2\tilde{l}_{\epsilon} r_q \lambda^{1/2}]}{8\lambda^{7/2} \left( r - \frac{\tilde{l}_{\epsilon} r_q}{\lambda^{1/2}}
\right)} + O(1)
\ena
Let us focus again on the case $\lambda=1$. The line element for $g_{\mu\nu}$ can now be conveniently expressed as

\be
ds^2=-\frac{A}{\Omega_{+}} dt^2 + \frac{\Omega_{+}}{A} \left(\frac{dx}{\Omega_{+}} \right)^2 + r^2(x) d\theta^2,
\en
where the radial coordinate $r(x)$ is explicitly written as

\be \label{eq:whcor}
r(x)=\frac{x  +\sqrt{x^2+4 r_q^2 \tilde{l}_{\epsilon}^2}}{2}.
\en

The behaviour close to the center, for any values of the parameters $M_0$ and $r_q$, and length scale $\tilde{l}_{\epsilon}^2$ is qualitatively the same. For
growing values of $\tilde{l}_{\epsilon}^2$ each curve smoothly deviates from their GR counterparts, and the finite-radius core beyond which they all cease to
exist grows as well. As in the GR case, there is always a cosmological horizon for any value of $\tilde{l}_{\epsilon}^2$ (see Fig.\ref{fig:4}).

Note that the reason for the existence of this finite-size core is due to the presence of the object $\Omega_{+}$ in the metric function $A/\Omega_{+}$, which
vanishes at $r=l_{\epsilon} r_q/\lambda^{1/2}$ when $l_{\epsilon}^2>0$, or at $r=\tilde{l}_{\epsilon} r_q/\lambda^{1/2}$ when $\tilde{l}_{\epsilon}^2>0$. The
latter also corresponds to $x=0$ due to the change of coordinates (\ref{eq:whcor}). Note that, in the equation defining the relation between coordinates,
$x^2=r^2 \Omega_{-}$, we have extracted a square root using the ($+$) solution. To extend the solution to the region $x<0$ in the case
$\tilde{l}_{\epsilon}^2>0$ we could consider the ($-$) solution there, thus replacing $x$ by its absolute value in (\ref{eq:whcor}). Bearing this in mind, the
solutions could be extended to the region $x<0$ using two charts, namely, $r_+$ for $x>0$ and $r_-$ for $x<0$ or, instead, a single chart for the coordinate $x$
covering the whole space-time, $x\in ]-\infty,+\infty[$. However, the matching of these geometries at $x=0$ is not smooth since $dr/dx \vert_{x=0} \neq 0$ [see
Eq.(\ref{eq:whcor})]. Indeed, at $x=0$, using Mathematica we find the Kretchsmann scalar, $K={R_\alpha}^{\beta\mu\nu}{R^\alpha}_{\beta\mu\nu}$, is dominated by
a curvature divergence of strength

\be
K \sim \log[x]^2/x^2,
\en
which is milder than in the GR case (for large distances we get $K \simeq 12r_q^2/r^4$, the standard GR behaviour). Despite this unpleasant behaviour, the
surface $x=0$ have some interesting properties. For example, an explicit computation of the electric charge flowing through the surface $x=0$ can be done
through the flux integral

\be
\Phi=\int_{S^1} *F=2\pi q,
\en
where $*F$ is Hodge dual, and the integration takes place over a boundary defined by close hypersphere $S^1$, where $2\pi$ is the volume of the unit $S^1$
sphere. An important property of this flux is that its rate of growth with the core ``area", $S=2\pi r$, defines a natural constant, i.e., the density of the
flux

\be
\frac{\Phi}{2\pi r} = \frac{\sqrt{\kappa^2/(4\pi)}}{\tilde{l}_{\epsilon}}
\en
is independent of the particular amounts of mass and charge, depending only on the fundamental constants $G$, $c$ (through Newton's gravitational constant in
three dimensions, $\kappa^2$) and the length scale $\tilde{l}_{\epsilon}$. When $\tilde{l}_{\epsilon}^2 \rightarrow 0$, this finite-structure disappears and the
standard point-like singularity of GR is recovered. However, the non-smoothness of the change of coordinates $dr/dx$ at $x=0$ seems to prevent a wormhole-type
extension, as opposed to the case of four dimensional Born-Infeld gravity \cite{or12}.

Let us now discuss briefly the case of solutions with $\lambda \neq 1$, according to the asymptotic behaviour given by Eq.(\ref{eq:asymp}). We have investigated
numerically this question and the following structures are found:

When $l_{\epsilon}^2>0$ and $\lambda>1$ we find the same number and type of horizons as in the GR case, but now with an asymptotically de Sitter behaviour,
instead of $\sim -2 r_q^2\log[r]$. For $l_{\epsilon}^2>0$ and $\lambda<1$ the behaviour is asymptotically Anti-de Sitter and we find that, as in the cases with
$\lambda=1$, a critical value $l_{\epsilon_c}^{2}(M_0,r_q,\lambda)$ determines a sudden transition. In this sense, those solutions with
$l_{\epsilon}^2<l_{\epsilon_c}^{2}$ correspond to naked singularities, while those with $l_{\epsilon}^2>l_{\epsilon_c}^{2}$ are black holes with a single
non-degenerate event horizon in all cases.

When $\tilde{l}_{\epsilon}^2>0$ we find asymptotically de Sitter solutions corresponding to $\lambda<1$, which have a cosmological horizon in all cases.
However, for $\lambda>1$ we find an asymptotically Anti-de Sitter behaviour and three classes of structures, depending on a critical value
$\tilde{l}_{\epsilon_c}^{2}(M_0,r_q,\lambda)$. In this sense, solutions with $\tilde{l}_{\epsilon}^2 > \tilde{l}_{\epsilon_c}^{2}$ correspond to black holes
with two horizons (inner and event), which merge into a extreme, degenerate horizon (extreme black holes) when
$\tilde{l}_{\epsilon}^2=\tilde{l}_{\epsilon_c}^2$, and solutions with $\tilde{l}_{\epsilon}^2<\tilde{l}_{\epsilon_c}^{2}$ correspond to naked singularities.

These behaviours resemble what it is usually find in other (Anti-)de Sitter backgrounds, like in the context of Gauss-Bonnet gravity \cite{GB}.

\section{Conclusions}

In this work we have considered two families of modified gravity theories sourced by electromagnetic fields. Using results from crystalline structures in solid
state physics, where the presence of point defects on their microstructure requires a metric-affine framework for their proper description at the macroscopic
level, we have obtained the equations for these theories from independent variations with respect to metric and connection. Motivated by the existence of some
ordered structures with defects which can be effectively treated as bi-dimensional systems (like graphene), we have considered metric-affine gravities with
non-metricity in a three dimensional space-time. The resulting formalism has several advantages with respect to the standard metric formalism of modified
gravity, like second-order equations in all cases or the recovery of GR in vacuum.

Explicit solutions to an $f(R)$ model were found and characterized. While for large distances they boil down to their GR counterparts, close to the center large
deviations are found. The case of $f(R)$ with $\alpha>0$ seems to be quite pathological, as the metric at the center of the solutions vanishes for any finite
value of $\alpha$, with a worse degree of divergence of the curvature invariants there. This has led us to consider a crystal-motivated action, as given by
Born-Infeld gravity, where a finite-size structure arises as long as $l_{\epsilon}^2$ is non-zero, replacing the standard point-like singularity of GR. When
$\lambda \neq 1$, and depending on the choice of parameters, standard horizon structures of black holes may arise. Our results support the interpretation of
$l_{\epsilon}^2$ as a natural scale around a point defect where the geometry may largely deviates from the GR counterpart. Further investigation on these
geometries could avoid the shortcoming of the typical space-time singularities of the Riemannian structure which, however, seems to be in conflict with the fact
that well defined geometries exist in the case of crystalline structures with defects. To get further into this question, one should consider the geodesic
behaviour in these scenarios with more realistic energy-matter sources, a topic which has received little attention in the literature so far. In summary, the
metric-affine formulation seems to be a more appropriate formalism to deal with defected crystals, which encourages further research on the field to identify
potential experimental signatures resulting from non-metricity. Progress in this sense is currently underway.

\section*{Acknowledgments}

This work was supported by the NSFC grants No.~11305038 and No.11450110403, the Shanghai Municipal Education Commission grant No.~14ZZ001, the Thousand Young Talents Program, and Fudan University. C.B. acknowledges also support from the Alexander von Humboldt Foundation. M.G.-N. acknowledges also support from China Scholarship Council (CSC), grant No.~2014GXZY08.



\end{document}